# Contactless transfer of angular momentum by rotating laser beam

E.V. Barmina and G.A. Shafeev

*Wave Research Center of A.M. Prokhorov General Physics Institute of the Russian Academy of Sciences, 38, Vavilov street, 119991 Moscow Russian Federation*

Abstract

Contactless transfer of angular momentum from rotating laser beam to a solid target is experimentally demonstrated. The effect is observed under irradiation of a glassy carbon target immersed in water by a pulsed laser beam that is scanned across the target surface along circular trajectory. The direction of target rotation coincides with that of the laser beam at small thickness of the liquid layer above the target while is opposite in case of higher thickness of the layer. The effect is interpreted as the interplay between thermocapillary and convective flows induced in the liquid by laser heating.
PACS: 47.20.Gv 47.32.Ef 47.57.J- 41.75.Jv 47.61.-k

Laser ablation of solids in liquid environment is a physical method for generation of nanoparticles of various elements and compounds (see for example, [1]). At sufficiently high laser fluence on the target the superficial layer melts, and the material is dispersed into surrounding liquid as nanoparticles. The generated nanoparticles leave the exposed area of the target due to strong convective and even turbulent flows induced by inhomogeneous heating of the liquid. Different methods are employed to increase the rate of nanoparticles generation, a flow cell reactor is one of them. Another method is scanning the laser beam with respect to the target. This can be achieved by electromagnetically deflected mirrors or by displacement of the target under stationary laser beam [2]. Laser radiation may induce bulk liquid motion owing to induced phase transition in a specially designed liquid due to light scattering [1]. Laser ablation of solids in liquids may also induce convective flows of the liquid due to self-organization of gas bubbles even in absence of the laser beam. This phenomenon is observed if the target was ablated according to a specially chosen pattern [3]. The collective self-organized flows emerge in the liquid due to interaction of numerous ascending bubbles of hydrogen between each other [4].

New interesting hydrodynamic phenomenon can be observed if the surface of a solid target fixed on the axis is ablated by scanning laser beam with closed trajectory. Namely, in certain experimental conditions the contactless transfer of angular momentum takes place from rotating laser beam to the target. In this communication we present the first experimental evidence of this effect in case of laser ablation of a glassy carbon target in water.

We used an yttrium-doped fiber laser with pulse duration of 200 ns and repetition rate of 20 kHz. Average power of the laser beam was 15 W at wavelength of 1.07 μm. Laser radiation was linearly polarized. Laser beam passed a system of two mirrors that could be deflected electromagnetically in X-Y directions. This allowed controlled displacement of the laser beam in the focal plane of the objective according to preset pattern. Linear velocity of the laser beam could be varied up to 4 m/s.

The target was made of a rounded plate of a glassy carbon. It was mounted on a needle allowing rotation of the target around its center. Mounted target was placed into a glass cell filled with water. Typical thickness of the liquid layer above the target surface was varied between 2 and 8 mm. It is limited by absorption of laser radiation in the colloidal solution of carbon nanoparticles, which decreases the level of laser intensity on the target.

Laser exposure of the target under water is accompanied by generation of carbon nanoparticles. These nanoparticles absorb laser radiation and promote ionization of the medium, which helps visualization of the laser beam inside the liquid. If the trajectory of the laser beam is a circle centered on the target center, then the target starts rotating around its axis. The picture taken during the laser exposure is presented in Fig. 1.

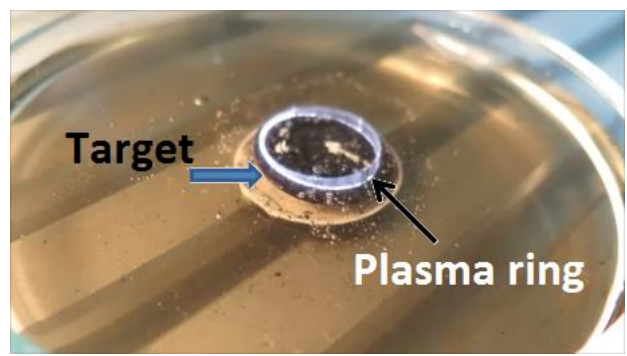

Fig. 1. Exposure of a glassy carbon target to rotating laser beam. Plasma ring is visible due to ionization of the medium on carbon nanoparticles. The beam rotates counter-clockwise with linear velocity of 2 m/s. The diameter of the target is 15 mm. White markers are drawn on the target to detect its rotation.

The target rotates in the same direction as the laser beam does. The angular velocity of target rotation depends on several experimental parameters, such as laser beam velocity, thickness of the liquid layer above the target, and average power of the laser beam. When the laser beam is switched off the rotation of the target stops in a fraction of a second [5]. Qualitatively similar behavior is observed if scanning is conducted along square-shaped trajectory, the directions of rotation of the beam and target remain the same. The angular velocity of the target increases with the decrease of the thickness of liquid layer.

The transfer of the momentum from rotating laser beam to the target is due to its viscous interaction of the flows induced above the target by laser heating. Two types of flows induced by laser beam can be distinguished. The first one is convective ascending flows with velocity $v_c$ that are due to heating of the liquid from the target in the field of gravity (Fig. 2). The second type of flows is due to the inhomogeneous heating of the surface of the liquid. Surface flows are directed from the center of the laser beam to the periphery of the cell due to Marangoni effect. In this particular case this surface flow is due to the decrease of the coefficient of surface tension σ of the liquid with temperature (so called thermocapillary flows). Typically $d\sigma/dT < 0$, and the liquid leaves the hot area and moves toward cold areas with velocity $v_s$. The temperature gradient is due to the absorption of laser radiation by nanoparticles of glassy carbon. Note that these thermocapillary flows exist even in absence of the target. The velocity $v_s$ has its maximal value at the border of laser beam on the liquid since temperature gradient in this region is the highest and gradually decreases towards the periphery of the cell due to viscosity of the liquid. Both flows are symmetrical in the case of heating with the stationary laser beam.

However, if the laser beam moves above the target at scanning velocity $v_{las} > v_s$, then the sum of the above mentioned flows is not symmetrical anymore.

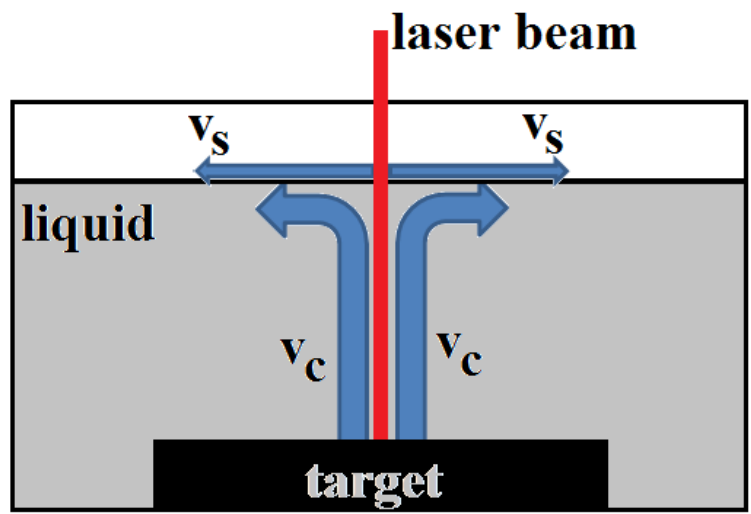

Fig. 2. Liquid flows in the cell under laser ablation of a solid target with stationary laser beam.

The ascending flow from the target surface moves with the laser beam. The surface flow, however, is not symmetrical anymore and its velocity is determined by $v_{las}$ rather than by $v_s$. The laser beam pushes the liquid along its trajectory. This creates a circular motion of the liquid above the target, which is then transferred to the target through viscous interaction. The laser beam acts here as a tea spoon. The video of rotating target pushed by laser beam can be found in the attached link [5]. No rotation of the target is observed if the laser exposure by rotating laser beam is conducted in air in the same range of laser power and scanning velocity $v_{las}$.

The angular velocity of the target rotation depends on the linear velocity $v_{las}$ of the laser beam. This dependence is presented in Fig. 3. The angular velocity was calculated by counting the number of $2\pi$ rotations of the target during filming. At high $v_{las}$ angular velocity of rotation tends to saturation. In this range of $v_{las}$ the angular velocity of target rotation is much smaller than that of the laser beam (2 s$^{-1}$ against 250 s$^{-1}$). At low $v_{las}$ (250 mm/s) rotation stops and is replaced by oscillations of the target for ±15º. With further decrease of $v_{las}$ the rotation recommences in the same direction.

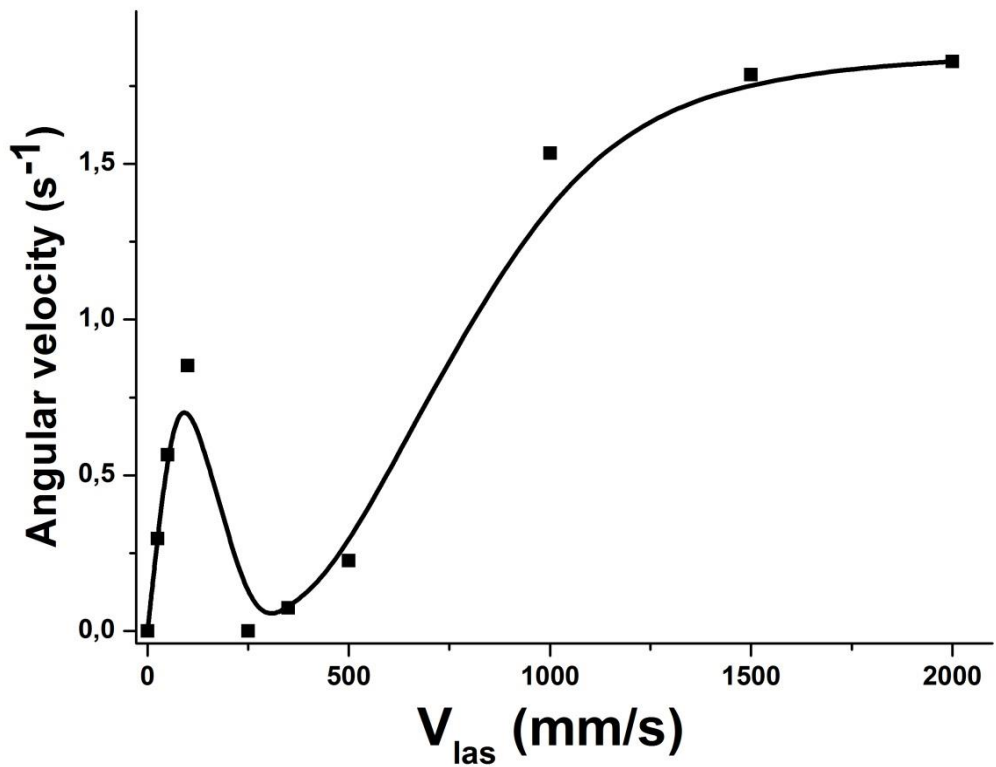

Fig. 3. Dependence of angular velocity of target rotation on scanning velocity of the laser beam. Thickness of the liquid above the target is of 4 mm.

Zero velocity of rotation observed at $v_{las}$ = 250 mm/s indicates to the change of direction of target rotation. The target oscillates back and forth at this combination of parameters, such as $v_{las}$ and thickness of the liquid layer. The direction of rotation changes its sign at higher thickness of the liquid layer. In this situation the direction of target rotation is opposite to that of the laser beam.

The reverse direction of rotation is due to the asymmetry of convective and surface flows of the liquid. At relatively high thickness of the liquid layer above the target the convective flows arrive to the surface of the liquid with some delay with respect to the position of the laser beam on the target. The resulting velocity of both convective and thermocapillary flows is directed in this case in the opposite direction with respect to direction of scanning.

Thus, a contactless transfer of angular momentum to a solid target has been experimentally demonstrated. The transmission proceeds via excitation of hydrodynamic flows in the liquid above the solid target. At relatively small thicknesses of the liquid layer about 2-4 mm the direction of target rotation coincides with that of the laser beam. At higher thicknesses, about $6 - 8$ mm, the direction of target rotation is opposite to that of the laser beam. Rotating laser beam induces rotating flows in the liquid, and the target serves their visualization. In fact, linear displacement of the two mirrors is converted into vortex flow of the liquid.

We thank A.V. Simakin, P.G. Kuzmin, A.A. Serkov, and I.I. Rakov for their help in conducting experiments.

References
1. *Laser ablation in liquids*, ed. by G.W. Yang, Pan Stanford, Singapore, 2011.
2. Robert D. Schroll, Régis Wunenburger, Alexis Casner, Wendy W. Zhang, and Jean-Pierre Delville, Liquid Transport due to Light Scattering, *Physical Review Letters*, **98**, 133601 (2007).
3. E.V. Barmina, P.G. Kuzmin, G.A. Shafeev, Self-organization of hydrogen bubbles rising above laser-etched metallic aluminum in a weakly basic aqueous solution, Physical Review E 84, 045302(R) (2011).
4. E. V. Barmina, N. A. Kirichenko, P. G. Kuzmin, and G. A. Shafeev, Self-organization of ascending-bubble ensembles, Physical Review E 87, 053001 (2013).
5. https://drive.google.com/file/d/0B1KzDrG_YKraRnFRZzA3aFREZ00/edit?usp=sharing